\documentclass[a4paper,fleqn,usenatbib,useAMS]{mnras}

\usepackage{array}
\usepackage{subfigure}
\usepackage{graphicx}	
\usepackage{amsmath}	
\usepackage{amssymb}	
\usepackage{multicol}        
\usepackage{bm}		
\usepackage{pdflscape}	

\usepackage[T1]{fontenc}
\usepackage{ae,aecompl}

\usepackage{newtxtext,newtxmath}

\title[Cosmic HI Density Distribution]
{The cosmic atomic hydrogen mass density as a function of mass and galaxy hierarchy from spectral stacking}

\author[Wenkai Hu et al.]
{Wenkai Hu$^{1,2,3,4}$\thanks{Contact e-mail: \href{wkhu@nao.cas.cn}{wkhu@nao.cas.cn}},
Barbara Catinella$^{2,4}$,
Luca Cortese$^{2,4}$,
Lister Staveley-Smith$^{2,4}$,
\newauthor
Claudia del P. Lagos$^{2,4}$,
Garima Chauhan$^{2,4}$,
Tom Oosterloo$^{6,7}$,
Xuelei Chen$^{1,3,5}$
\\
$^{1}$ Key Laboratory for Computational Astrophysics, National Astronomical Observatories, Chinese Academy of Sciences, 20A Datun Road, Beijing 100012, China
\\
$^{2}$ International Centre for Radio Astronomy Research (ICRAR), M468, University of Western Australia, 35 Stirling Hwy, WA 6009, Australia\\
$^{3}$ School of Astronomy and Space Science, University of Chinese Academy of Sciences, Beijing 100049, China\\
$^{4}$ ARC Centre of Excellence for All Sky Astrophysics in 3 Dimensions (ASTRO 3D)\\
$^{5}$ Center of High Energy Physics, Peking University, Beijing 100871, China\\
$^{6}$ ASTRON, the Netherlands Institute for Radio Astronomy, Postbus 2, 7990 AA, Dwingeloo, The Netherlands\\
$^{7}$ Kapteyn Astronomical Institute, University of Groningen, P.O. Box 800, 9700 AV Groningen, The Netherlands\\
}

\date{Last updated 2015 May 22; in original form 2013 September 5}

\pubyear{2019}

\begin{document}
\label{firstpage}
\pagerange{\pageref{firstpage}--\pageref{lastpage}}
\maketitle

\begin{abstract}
We use spectral stacking to measure the contribution of galaxies of different masses and in different hierarchies to the cosmic atomic hydrogen (HI) mass density in the local Universe. Our sample includes 1793 galaxies at $z < 0.11$ observed with the Westerbork Synthesis Radio Telescope, for which Sloan Digital Sky Survey spectroscopy and hierarchy information are also available.
We find a cosmic HI mass density of $\Omega_{\rm HI} = (3.99 \pm 0.54)\times 10^{-4} h_{70}^{-1}$ at $\langle z\rangle = 0.065$. For the central and satellite galaxies, we obtain $\Omega_{\rm HI}$ of $(3.51 \pm 0.49)\times 10^{-4} h_{70}^{-1}$ and $(0.90 \pm 0.16)\times 10^{-4} h_{70}^{-1}$, respectively. 
We show that galaxies above and below stellar masses of $\sim$10$^{9.3}$ M$_{\odot}$ contribute in roughly equal measure to the global value of $\Omega_{\rm HI}$. While consistent with estimates based on targeted HI surveys, our results are in tension with previous theoretical work. We show that these differences are, at least partly, due to the empirical recipe used to set the partition between atomic and molecular hydrogen in semi-analytical models. Moreover,  comparing our measurements with the cosmological semi-analytic models of galaxy formation {\sc Shark} and GALFORM reveals gradual stripping of gas via ram pressure works better to fully reproduce the properties of satellite galaxies in our sample, than strangulation. Our findings highlight the power of this approach in constraining theoretical models, and confirm the non-negligible contribution of massive galaxies to the HI mass budget of the local Universe.
\end{abstract}




\section{Introduction}
Neutral atomic hydrogen (HI) plays a key role in the formation and evolution of galaxies. As the simplest, most abundant, and spatially extended galactic gas component, atomic hydrogen is important to understand a wide range of astrophysical processes such as star formation histories and galaxy interactions, as well as trace the cosmic large$-$scale structure. 

In recent years, observational constraints on the HI content of galaxies have become available for local and higher-redshift samples. The HI Parkes All-Sky Survey  \citep[HIPASS;][]{2001MNRAS.322..486B} has detected HI emission from 5,317 galaxies at $0 < z < 0.04$ over a sky area of 21,341 deg$^{2}$ \citep{2004MNRAS.350.1195M,2006MNRAS.371.1855W}, and the Arecibo Legacy Fast ALFA (ALFALFA) survey \citep{2005AJ....130.2598G} has detected $\sim$ 31,500 galaxies out to $z = 0.06$ over a sky area of approximately 7,000 deg$^{2}$ \citep{2018ApJ...861...49H}. These large-area surveys allow for accurate measurement of the local HI mass function and the cosmic HI gas density \citep{2005MNRAS.359L..30Z,2010ApJ...723.1359M,2018MNRAS.tmp..502J}. 

Beyond the local Universe, HI emission has been detected from galaxies up to $z \sim 0.3$ with deep integrations \citep{2001Sci...293.1800Z,1538-4357-668-1-L9,2008ApJ...685L..13C,2015MNRAS.446.3526C,2016ApJ...824L...1F}. The ongoing COSMOS HI Large Extragalactic Survey (CHILES)
with the upgraded Jansky Very Large Array is imaging HI over the $z=0-0.45$ redshift interval and holds the current record for the highest-redshift HI emission detection at $z=0.376$ \citep{2016ApJ...824L...1F}.




At the same time, studies of the HI gas content of galaxies in different environments reveal that galaxies in dense regions are usually HI deficient 
\citep{1973MNRAS.165..231D,1984ARA&A..22..445H,2001ApJ...548...97S,2011MNRAS.415.1797C,2013MNRAS.436...34C,2016ApJ...824..110O,2016ApJ...832..126S,2017MNRAS.466.1275B}, whereas gas-rich galaxies are typically found in the most weakly clustered regions \citep{2007ApJ...654..702M,2012ApJ...750...38M}.  


In addition to direct HI detection, the spectral stacking technique has also been successfully used to probe HI in galaxies out to $z \sim 1.45$ \citep{1538-4357-668-1-L9,2009MNRAS.399.1447L,2018ApJ...865...39B} and to quantify gas scaling relations of nearby galaxies \citep{2011MNRAS.411..993F} and their dependence on environment and active galactic nuclear activity \citep{2012MNRAS.427.2841F,2015MNRAS.452.2479B,2018MNRAS.473.1868B,2011arXiv1104.0414F,2013A&A...558A..54G,2019ApJ...882L...7B}. The cosmic HI gas density has also been successfully constrained at different redshifts (0.0$-$0.37) \citep{2007MNRAS.376.1357L,2013MNRAS.433.1398D,2013MNRAS.435.2693R,2016MNRAS.460.2675R,2018MNRAS.473.1879R}. In particular, \citet{2016ApJ...818L..28K} used the Giant Metrewave Radio Telescope (GMRT) to stack HI emission from massive star-forming galaxies at $z\sim 1.18 –- 1.34$, the highest redshift measurement of HI flux ever made using HI spectral stacking.

Despite these successes and the general agreement on the estimate of the global HI mass density in the local Universe, the relative contribution of different types of galaxies to $\Omega(\rm HI)$ is still under debate. \citet{2010MNRAS.408..919S} measured the cumulative HI mass density above a given HI mass for $\sim$190 galaxies with $M_{\ast}>10^{10}M_{\odot}$, obtained from the GALEX Arecibo SDSS Survey \citep[GASS;][]{2010MNRAS.403..683C}. They found that 36 $\pm$ 5 per cent of the total HI mass density is from galaxies with $M_{\ast}>10^{10}M_{\odot}$. \citet{2013ApJ...776...74L} presented the bivariate atomic hydrogen--stellar mass function for 480 galaxies in the GASS Data Release 2 \citep{2012A&A...544A..65C}, finding that massive systems ($M_{\ast}>10^{10}M_{\odot}$) contribute 41 per cent of the HI density in the local Universe. These results from observations consistently show that a significant fraction of the HI mass in the local Universe is associated with massive galaxies. However, these findings appear in contradiction with what found in cosmological simulations. Using the GALFORM model of galaxy formation set in the cold dark matter ($\Lambda$CDM) framework, \citet{2014MNRAS.440..920L} studied the contribution of galaxies with different properties to the global HI density. They predicted that the density of HI is always dominated by galaxies with low stellar masses ($M_{\ast} < 10^9 M_{\odot}$) and only $\sim$ 9 per cent of the HI density is contributed by galaxies with $M_{\ast}>10^{10}M_{\odot}$. The difference between observations and simulations suggests that the mechanisms driving the HI distribution in galaxies with different stellar masses are not yet well understood. Whether this is an issue with current data or a limitation of state-of-the-art numerical models is still unclear. 

In order to help solving this apparent tension between theory and observations, in this paper we quantify the contribution of galaxies of different masses to the cosmic HI density. We compare our results with previous observations and simulations and try to explain the differences emerging from previous works. Moreover, we quantify for the first time the contribution of centrals and satellites to $\Omega(\rm HI)$.

In \citet[][hereafter Paper I]{10.1093/mnras/stz2038}, we developed an interferometric stacking technique to study the HI content of galaxies at $z<0.12$, yielding an accurate measurement of the cosmic HI density in the local Universe and confirming that there is little evolution in $\Omega_{\rm HI}$ at low redshift. In this paper we use the same sample and technique to further explore the contribution of centrals and satellites to the cosmic HI density.

This paper is organized as follows: Section~\ref{sec:sample} describes the observational data and the galaxy group catalogue used in this work. We summarize the spectral extraction and stacking methodology in Section~\ref{sec:script}, present our measurements of $\Omega_{\rm HI}$ as a function of stellar mass and hierarchy in Section~\ref{sec:density}, and compare these with semi-analytic model simulations in Section~\ref{sec:simulation}. In Section~\ref{sec:discussion} we discuss the implications of our results for our understanding of the gas cycle in galaxies. Throughout this paper we use H$_{\circ} = 70$ km s$^{-1}$ Mpc$^{-1}$, $\Omega_{\rm m} = 0.3$ and $\Omega_{\Lambda} = 0.7$.

\section{Sample}
\label{sec:sample}
\subsection{HI Data}
The sample used in this work is described in detail in Paper I. Briefly, the
HI observations were carried out with the Westerbork Synthesis Radio Telescope (WSRT), and consisted of 36 individual pointings in a strip of the Sloan Digital Sky Survey \citep[SDSS;][]{2000AJ....120.1579Y} South Galactic Cap (21h < RA < 2h and $10^{\circ}$< DEC <$16^{\circ}$. Each pointing was observed with an integration time varying between 5 hr and 12 hr, for a total observing time of 351 hr. Data from one of the pointings were discarded due to bad quality. The half-power beam width (HPBW) of WSRT is 35$\arcmin$, and the average synthesized beam size is $108\arcsec \times 22\arcsec$. The overall frequency range for the reduced data is 1.406 GHz to 1.268 GHz, corresponding to a redshift range of 0.01 < z < 0.12. However, due to stronger radio frequency interference (RFI) at higher redshift we set an upper redshift limit of z = 0.11.

\subsection{Optical data}
We use SDSS Data Release 7 \citep{2009ApJS..182..543A} as the optical catalogue for our stacking analysis. With the target selection algorithm described in \citet{2002AJ....124.1810S}, the SDSS sample has a completeness that exceeds 99$\%$ (excluding fibre collisions). We extract SDSS spectroscopic targets within the footprint of our WSRT observations. This is defined by the regions where the normalized primary beam response is above 0.1. This provides us with a sample of 1,895 galaxies spanning the redshift range $0.01 < z < 0.11$. We complement the photometric information provided by the SDSS catalog with stellar masses taken from the MPA-JHU (Max-Planck Institute for Astrophysics - John Hopkins University) value-added galaxy catalogue \citep{2003MNRAS.341...33K}.

\subsection{Galaxy Group Catalogue}
In order to identify centrals and satellite galaxies in our sample, we use a dark matter halo group catalogue based on the galaxies in the SDSS main galaxy sample with redshift completeness C $\geq$ 0.7 \citep{2007ApJ...671..153Y,2012ApJ...752...41Y}. The first Yang group catalogue derived from the SDSS DR4 (\citealt{2007ApJ...671..153Y}) used about 362,356 galaxies to identify groups in the redshift range 0.01 < z < 0.2. Extending their analysis to SDSS DR7 (\citealt{2012ApJ...752...41Y}) they increased the number of galaxies to $\sim$ 599,300.
In this catalogue the dark matter halos are identified using the following iterative process: (1) identification of potential group centres; (2) calculation of the group luminosity for each tentative group; 
(3) estimation of mass, size and velocity dispersion of the dark matter halo associated with it (initially using a constant mass-to-light ratio for all groups); (4) based on the properties of the associated halo, the candidate group members might be reassigned; (5) a new group centre is then computed and the process is iterated until there is no further change in the group membership. The final halo masses are assigned via abundance matching, using the halo mass function derived by \citet{2006ApJ...646..881W}.

Here, we adopt the assumption that the galaxy with the largest stellar mass is the central galaxy. Other galaxies in the group will be called satellites. 

\begin{figure}
    \centering
    \includegraphics[width=8cm]{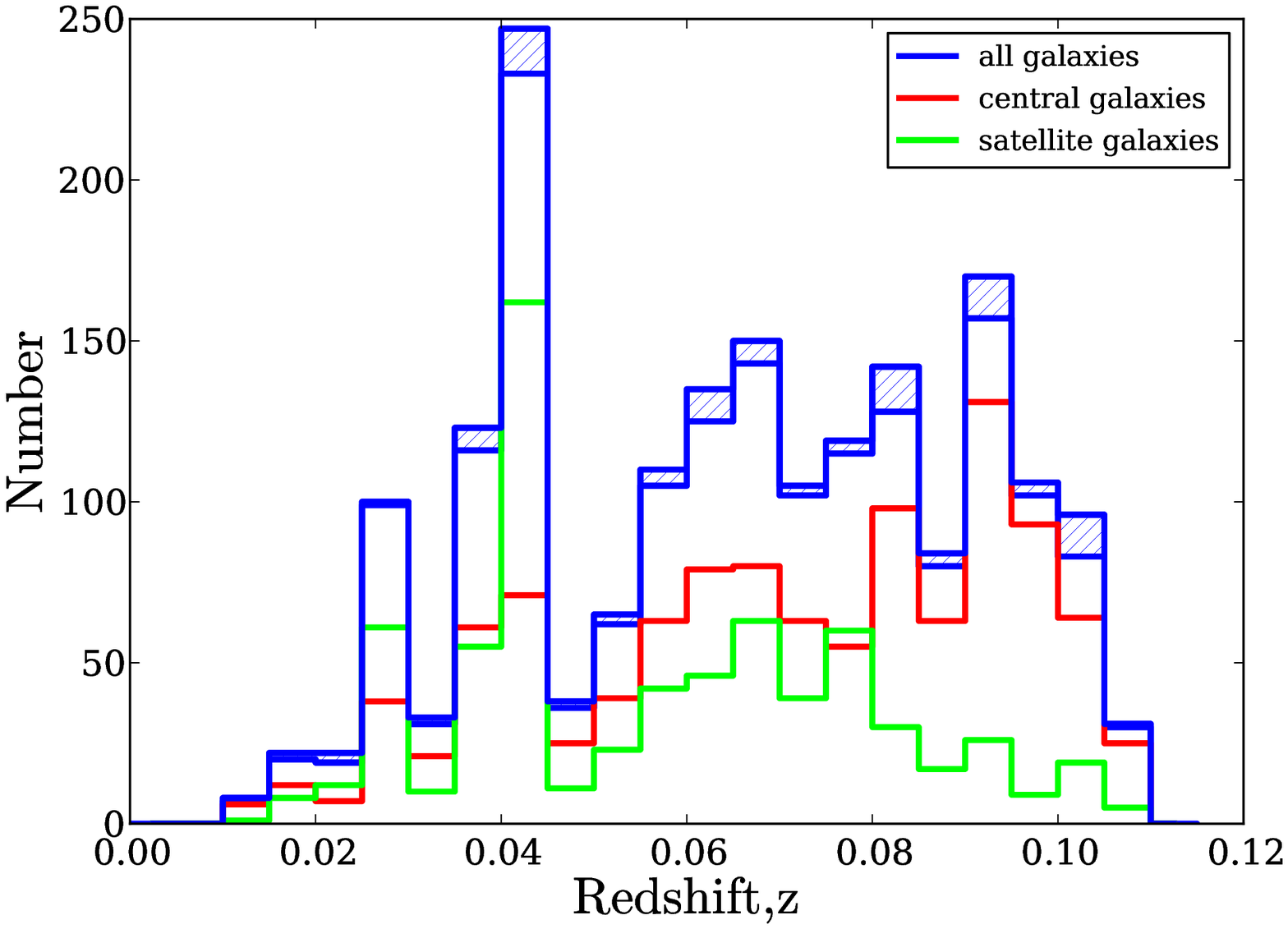}
    \includegraphics[width=8cm]{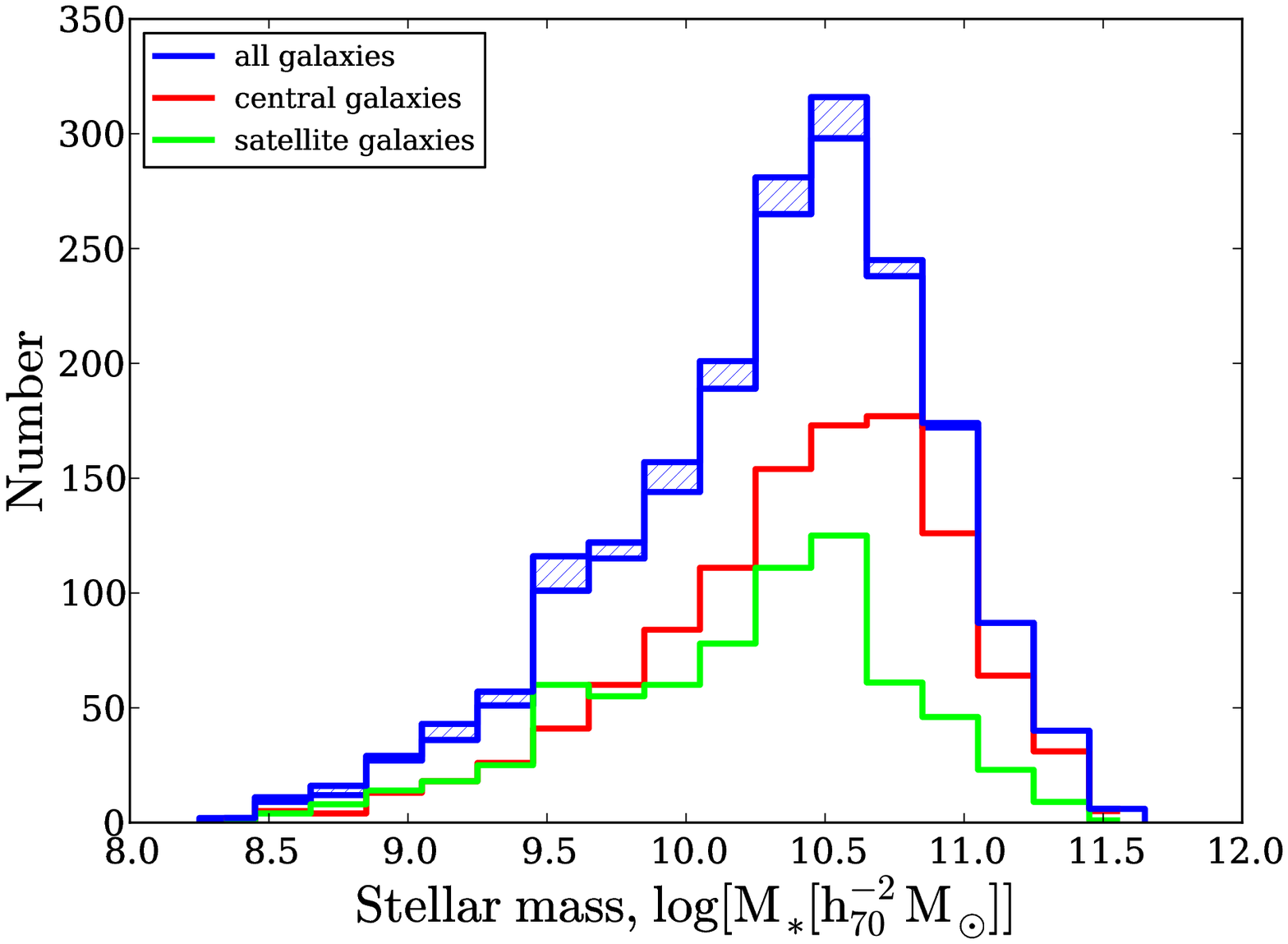}
    \caption{Redshift (top panel) and stellar mass (bottom panel) distributions of the SDSS spectroscopic subsample contained within our 35 WSRT pointings (blue). The distributions of central and satellite galaxies are shown by red and green lines, respectively. The hatched blue regions show the missing galaxies after cross-matching with the Yang catalogue (see text). 
    The missing galaxies tend to be faint and low-mass systems. The intervals for redshift and stellar mass bins are 0.005 and 0.2 dex. 
    }
    \label{distribution}
\end{figure}

The group catalogue excludes galaxies with a redshift completeness C $<$ 0.7. The cross-matching of the source list of our pointings (SDSS DR7) and the Yang catalogue (DR7) reduces the number of galaxies in our sample by $\sim 5\%$, from 1895 to 1793 galaxies; the matched subset has a mean redshift of $\langle z\rangle$ = 0.065. We show the redshift and stellar mass distribution of central (red histogram), satellite (green) and all galaxies (blue) in Figure~\ref{distribution}. 
Of the 1793 galaxies, 699 (39$\%$) are classified as satellites and 1094 (61$\%$) as centrals, of which 906 are isolated. In what follows, we consider the isolated galaxies as central galaxies. We note that 350 galaxies do not have associated halo masses, as the group catalogue does not assign halo masses to very small halos and/or isolated centrals with low stellar mass. However, this does not affect our analysis, which is based only on the central/satellite distinction.

\section{Stacking Procedure}
\label{sec:script}
The stacking technique used in this paper is described in detail in Paper I. In summary, after the removal of residual continuum emission from very bright sources, the HI spectra are de-redshifted and the HI flux density is conserved by applying: $S_{\nu_{\rm res}} = S_{\nu_{\rm obs}}/(1+z)$. The flux spectra are converted into mass spectra using the following relation:
\begin{eqnarray}
    m_{\rm HI}(\nu) = 4.98\times10^{7} S_{\nu} D_{L}^{2}f^{-1} ,
    \label{mass_spectrum}
\end{eqnarray}
where $S_{\nu}$ is the de-redshifted HI flux density in Jy, $D_{L}$ is the luminosity distance in Mpc, $f$ is the normalized primary beam response, and $m_{\rm HI}$ is in units of M$_{\sun}$ MHz$^{-1}$. The spectrum of $i$-th galaxy is weighted by:
\begin{eqnarray}
    w_{i} = f^{2}D_{L}^{-1}\sigma^{-2},
    \label{weight}
\end{eqnarray}
where $\sigma$ is the rms noise of the flux density spectrum in Jy. The averaged final stacked spectrum is obtained from:
\begin{eqnarray}
    \langle m_{\rm HI}(\nu)\rangle = \frac{\sum_{i=1}^{n}w_{i}m_{\rm HI,i}}{\sum_{i=1}^{n}w_{i}}.
    \label{weighted_spectrum}
\end{eqnarray}
The integrated HI mass of a stack is then defined as the integral along the frequency axis over the mass spectrum:
\begin{eqnarray}
    \langle M_{\rm HI}\rangle = \int_{-\Delta\nu}^{\Delta\nu}\langle m_{\rm HI}(\nu)\rangle d\nu,
    \label{integrated_mass}
\end{eqnarray}
where $\Delta\nu = 1.5$ MHz in this paper, corresponding to $\pm 317$ km s$^{-1}$. 

We estimate the error on the HI mass measurement through jackknife resampling. From the total sample of $n$ spectra, $n/20$ randomly selected spectra are removed at a time to construct 20 jackknife samples, from which 20 mass spectra are obtained. The jackknife estimate of the true variance of the measured value of the integrated HI mass is then given by:
\begin{eqnarray}
     \sigma^2(\langle M_{\rm HI}\rangle) = \frac{19}{20}\sum_{j=1}^{20}(\langle M_{\rm HI}\rangle-\langle M_{\rm HI}\rangle^{j}\rangle)^{2},
    \label{jackknife}
\end{eqnarray}
where $\langle{M}_{\rm HI}\rangle$ refers to the averaged HI mass spectrum from the original sample.
We can also measure $\langle M_{\rm HI}/L\rangle$ and its error by stacking the individual $M_{\rm HI}/L$ spectra. We do this via Equation~\ref{weighted_spectrum} and \ref{integrated_mass}, with $M_{\rm HI}$ replaced by $M_{\rm HI}/L$.

\subsection{Confusion Correction}
The value of the average HI mass measured via this stacking method is potentially increased by beam confusion. In other words,
individual spectra might be contaminated by additional HI flux from neighbouring galaxies at similar recessional velocity as the
targeted galaxy, located within the WSRT beam and spectra extraction region.
Although the WSRT synthesized beam is small, $\sim 7$ per cent of our sample is potentially confused with neighbouring galaxies. We follow the method in \citet{2012MNRAS.427.2841F} to model the confusion, estimating the total signal $S_{i}$ as the sum of the sample galaxy $S_{s}$ and the companions ($S_{c}$) weighted with two factors:
\begin{eqnarray}
     S_{i} = S_{s} + \Sigma_{c}f_{1;c}f_{2:c}S_{c},
     \label{signal_with_confusion}
\end{eqnarray}
where $f_{1}$ and $f_{2}$ model the overlap between the sample galaxy and its companion in angular and redshift space. These are given by:
\begin{eqnarray}
     f_{1} =  \exp[-0.5\times(\frac{x}{\sigma_{x}})^2 - 0.5\times(\frac{y}{\sigma_{y}})^2],
    \label{f1}
\end{eqnarray}
\begin{eqnarray}
     f_{2} = \delta w/w_{s},
    \label{f2}
\end{eqnarray}
where x, y are the projected angular distances between sample galaxies and the companion, $\sigma_{x} = (2\sqrt{2\rm ln2})\times22$ arcsec and $\sigma_{y} = (2\sqrt{2\rm ln2})\times108$ arcsec. $w$ is the expected HI line width and $\delta w$ is the velocity overlap between the sample galaxy $S_{s}$ and the companion. We evaluate the expected width by: $w_{obs} = w_{TF}\sin(i)$, and $w_{TF}$ is estimated from the r-band Tully-Fisher relation from \citet{2007AJ....134..945P}. The inclination $i$ is given by \citep{2007ApJS..172..599S}:
\begin{eqnarray}
     \rm (\cos incl)^{2} = \frac{(b/a)^2-(b/a)^2_{eos}}{1-(b/a)^2_{eos}},
     \label{inclination}
\end{eqnarray}
where $b/a$ is the r-band disk axis ratio from the SDSS catalogue (a and b are the semimajor and semiminor axis, respectively), and $(b/a)_{eos} = 0.2$ is the intrinsic axial ratio for an edge-on spiral \citep{2007ApJS..172..599S}.

Finally, the expected HI mass of each companion is estimated using the relation between $M_{\rm HI}$ and galaxy optical diameter \citep{2011ApJ...732...93T}:
\begin{eqnarray}
     \log(M_{\rm HI}/M_{\odot}) = 8.72 + 1.25\log(D_{25,r}/kpc),
     \label{expected_HI_mass}
\end{eqnarray}
where the r-band diameter, $D_{25,r}$, is calculated following \cite{2016ApJ...824..110O} as:
\begin{eqnarray}
     \log D_{25,r} = \log(isoA_{r} ~\rm 0.39^{\prime\prime} ~adist) + 0.35\log(b/a),
     \label{diameter}
\end{eqnarray}
where $isoA_{r}$ is the r-band isophotal major axis in pixels, $0.39^{\prime\prime}$ arcsec$^{-1}$ is the SDSS pixel scale, and $adist$ is the number of kiloparsecs per arcsecond at the distance of the galaxy. 

With all the parameters given above, the true signal from the sample galaxy is:
\begin{eqnarray}
     S_{s} = S_{i} - \Sigma_{c}f_{1;c}f_{2:c}S_{c},
     \label{signal_without_confusion}
\end{eqnarray}
The confusion correction will be applied later to all the stacking measurements of real data. However, we find that this correction is pretty small for our sample -- the uncorrected values of $\langle M_{\rm HI}\rangle$ for all galaxies, satellites only and centrals only are 1.4, 1.8 and 1.4 per cent larger than the corresponding results obtained after applying our confusion correction, respectively (see also Paper I).

\section{Cosmic HI Density as a function of stellar mass and hierarchy}
\label{sec:density}
\subsection{Splitting centrals and satellites}
In principle, the cosmic HI density $\rho_{\rm HI}$ can be computed as:
\begin{eqnarray}
     \rho_{\rm HI} = \int M_{\rm HI}(M_{\ast})\phi_{M_{\ast}}(M_{\ast})~dM_{\ast},
    \label{hidensity_mstellar_int}
\end{eqnarray}
where $\phi_{M_\ast}(M_\ast)$ is the stellar mass function. For consistency with the group catalog used here, we adopt the stellar mass function estimate by \citet{2009ApJ...695..900Y}, based on 369,447 SDSS galaxies with redshifts in the range 0.01 $\leqslant$ z $\leqslant$ 0.20 
and parameterised as a Schechter function \citet{1976ApJ...203..297S}:
\begin{eqnarray}
     \phi_{M_{\ast}}(M_{\ast})dM_{\ast} = \phi_{M_{\ast}}^{\star}\left(\frac{M_{\ast}}{M^{\star}}\right)^{\alpha}\exp\left(-\frac{M_{\ast}}{M^{\star}}\right)\frac{dM_{\ast}}{M^{\star}}.
    \label{schechter_stellar}
\end{eqnarray}
The normalization $\phi_{M_{\ast}}^{\ast}$, turnover point $M^{\ast}$ and low-mass end slope $\alpha$ for all, satellite and central galaxies are listed in Table~\ref{smf_parameters}. The HI density in each stellar mass bin can then be obtained as:
\begin{eqnarray}
     \rho_{\rm HI}(M_{\ast}^{i})\Delta M_{\ast}^{i} = \langle M_{\rm HI}(M_{\ast}^{i})\rangle\times\phi_{M_{\ast}}(M_{\ast}^{i})\times\Delta M_{\ast}^{i},
    \label{hidensity_mstellarbins}
\end{eqnarray}
once $\langle M_{\rm HI}(M_{\ast}^{i})\rangle$ is estimated by stacking galaxies in our sample per bin of stellar mass. 

\begin{figure}
    \centering
    \includegraphics[width=8cm]{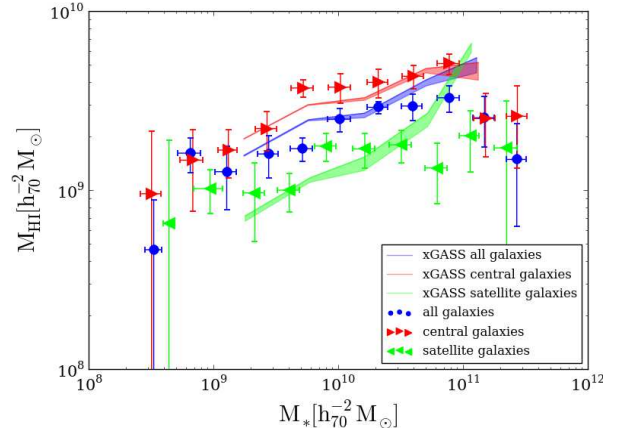}
    \caption{Stacking all galaxies (blue symbols), centrals only (red) and satellites only (green) in stellar mass bins shows that the relation between $\langle M_{\rm HI}\rangle$ and $M_{\ast}$ cannot be modeled by a simple power law. For comparison, we show the results obtained using the xGASS representative sample (\citealp{2018MNRAS.476..875C}, coloured lines).}
    \label{stacking_stellarmassbins}
\end{figure}

The result of this stacking procedure is shown in Figure~\ref{stacking_stellarmassbins}. We recover the well known increase of atomic gas mass with stellar mass, and confirm that central galaxies (red) have significantly larger HI reservoirs than satellites (green) at all stellar masses. We compare our results with those obtained from the extended GASS survey \citep[xGASS;][]{2018MNRAS.476..875C}, a targeted and {\rm HI} gas-fraction-limited survey of 1179 galaxies selected only by stellar mass (10$^{9}$M$_{\odot}$<M$_{\ast}$<10$^{11.5}$M$_{\odot}$) and redshift (0.01<z<0.05). We use the xGASS representative sample, excluding galaxies flagged as confused, and estimate average HI masses per bin of stellar mass using Eq.~\ref{integrated_mass}. Given that xGASS  includes non-detections, we estimate the average HI content in two ways, by setting the {\rm HI} masses of the non-detections to their upper limits or to zero. The difference (generally negligible) between the two approaches is shown by the thickness of the lines in Fig.~\ref{stacking_stellarmassbins}. We find that the stacking of our WSRT data produces results consistent with those obtained from xGASS. This is not trivial, as observations and techniques are significantly different. The only tension is for the most massive (i.e., stellar masses $>$10$^{10.5}$ M$_{\odot}$) satellites, for which our stacking technique predicts HI masses a factor of $\sim$2 lower than xGASS. This is likely due to the different selection of the two samples but, as we show below, it does not affect our results. Indeed, this would only strengthen our main conclusion that massive galaxies significantly contribute to the cosmic HI density in the local Universe.



The overall agreement between our stacking procedure and xGASS gives us confidence on the reliability of our approach. We can thus take advantage of Eq.~\ref{hidensity_mstellarbins} to estimate how the HI mass density in galaxies varies as a function of stellar mass. As shown in Fig.~\ref{rho_HI_function_stellarmass}, the distribution of HI density as a function of stellar mass is well approximated by a Schechter function (see Table~\ref{density_function_parameters}), with the knee of the distribution clearly above 10$^{10}$ M$_{\odot}$ and an either declining or flat slope at low stellar masses. By integrating the fitted Schechter functions for all galaxies, and for centrals and satellites separately, we find the following values of cosmic HI density in the local Universe: 
\begin{eqnarray}
     \Omega_{\rm HI} = (3.99 \pm 0.54)\times 10^{-4}h_{70}^{-1},\\
     \Omega_{\rm HI,ce} = (3.51 \pm 0.49)\times 10^{-4}h_{70}^{-1},\\
     \Omega_{\rm HI,sa} = (0.90 \pm 0.16)\times 10^{-4}h_{70}^{-1},
     \label{omega_HI_group}
\end{eqnarray}
where the error is estimated with error propagation. For the integrations here and below, we do not include the HI mass in galaxies with $M_{\ast} < 10^{5}M_{\odot}$.
Our value of $\Omega_{\rm HI}$ is consistent with that presented in Paper I (($4.02 \pm 0.26) \times 10^{-4}h_{70}^{-1}$), although based on a slightly restricted sample and different technique, as well as with previous literature values determined using either HI stacking \citep{2013MNRAS.433.1398D,2007MNRAS.376.1357L,2016ApJ...818L..28K,2013MNRAS.435.2693R,2016MNRAS.460.2675R,2018MNRAS.473.1879R} or 21-cm emission detections \citep{2005MNRAS.359L..30Z,2010ApJ...723.1359M,2011ApJ...727...40F,2015MNRAS.452.3726H,2018MNRAS.tmp..502J}.

The sum of the HI densities of central and satellite galaxies, $\Omega_{\rm HI,ce}$ and $\Omega_{\rm HI,sa}$, is $(4.41 \pm 0.52) \times 10^{-4}h_{70}^{-1}$, which is consistent with the result from the measurement using all galaxies, confirming that our technique is self-consistent. About $\sim$80\% percent of the HI content is located in central galaxies, with satellites contributing less than $\sim$20\%. Combining with the stacking results, we find that at low redshift central galaxies not only have larger average HI masses than satellite galaxies, but also contain most of the HI content in the Universe. Of course, this is entirely expected: not only centrals dominate satellites in numbers at all stellar masses, but satellite galaxies are also generally gas-poorer than centrals at fixed stellar mass \citep[e.g.,][]{2013MNRAS.436...34C}.

\begin{table}
 \caption{The parameters of the stellar mass functions \citep{2009ApJ...695..900Y} used in this paper.}
 \centering
 \label{smf_parameters}
  \begin{tabular}{lccc}
   \hline
    Populations & $\phi^{\ast}_{M_{\ast}}$ & $\alpha$ & log$M^{\ast}$\\
     & (Mpc$^{-3}$d$M_{\ast}$) & & (M$_{\odot}$)\\
   \hline
    All     & 2.30$\times10^{-3}$ & $-$1.16 & 11.03\\
    satellites & 1.03$\times10^{-3}$ & $-$1.08 & 10.79\\
    centrals   & 1.62$\times10^{-3}$ & $-$1.14 & 11.07\\
   \hline
  \end{tabular}
\end{table}

Before we proceed, it is important to note that the technique used to estimate $\Omega_{\rm HI}$ in this work is significantly different from that using the $\langle M_{\rm HI}/L\rangle$ bias correction presented in Paper I, where we first estimated the mean HI mass-to-light ratio of galaxies via stacking and then bootstrapped from the SDSS luminosity function. As SDSS is magnitude-limited, many optically faint but HI-rich galaxies are missing. Thus, to correct for this selection bias, in Paper I we derived a weight factor (C1) that accounts for the different mass-to-light ratios of the sample compared to an unbiased selection of galaxies. Using this method and the luminosity functions for all, satellites and centrals given by \citet{2009ApJ...695..900Y}, we find:
\begin{eqnarray}
     \Omega_{\rm HI} = \frac{\rho_{\rm HI}}{\rho_{\rm c,0}} = (4.26 \pm 0.36)\times 10^{-4}h_{70}^{-1}.
     \label{omega_HI_C1_all}
\end{eqnarray}
This value of $\Omega_{\rm HI}$ is consistent with the result in Paper I ($\Omega_{\rm HI} = (4.02 \pm 0.26)\times10^{-4}h_{70}^{-1}$), although the sample in this work is not exactly the same (we lost galaxies that have no matches in the group catalog). For the centrals and satellites the same technique provides (see Table ~\ref{omegahi_group_C1}):
\begin{eqnarray}
     \Omega_{\rm HI,ce} = (3.53 \pm 0.37)\times 10^{-4}h_{70}^{-1},\\
     \Omega_{\rm HI,sa} = (0.96 \pm 0.14)\times 10^{-4}h_{70}^{-1}.
     \label{omega_HI_C1_group}
\end{eqnarray}
These values are consistent with the measurements presented above, and derived from the Schechter function fitting to the stacking in stellar mass bins, suggesting that the two methods are self-consistent. 

While in principle we could have directly measured $\langle M_{\rm HI}/M_{\ast}\rangle$ and used a correction factor to compute the HI density as we did in Paper I, in practice this is less robust. Indeed, while $\langle M_{\rm HI}/L\rangle$ at low r-band luminosity can be extrapolated using a power-law relation between $\langle M_{\rm HI}/L\rangle$ and luminosity, there is no simple relation between $\langle M_{\rm HI}/M_{\ast}\rangle$ and $M_{\ast}$ \citep{2018MNRAS.481.3573L,2018MNRAS.476..875C,2018ApJ...864...40P}.

\begin{table*}
	\centering
	\caption{Measurement of $\Omega_{\rm HI}$ for all galaxies, satellites only and centrals only.}
	\label{omegahi_group_C1}
	\begin{tabular}{lccccc} 
		\hline
		Populations & Number of galaxies & C1 & $\langle M_{\rm HI}/L_{r}\rangle$ & $\rho_{L}$ & $\Omega_{\rm HI}$\\
		 & & & M$_{\odot}/$L$_{\odot}$& $(10^8 h_{70}$ L$_{\odot}$ Mpc$^{-3})$ & $(10^{-4}h_{70}^{-1})$\\
		\hline
		All        & 1793 & 1.56 & 0.29 $\pm$ 0.02 & 1.30 & 4.26 $\pm$ 0.36\\
        satellites & 699 & 2.02 & 0.21 $\pm$ 0.03 & 0.31 & 0.96 $\pm$ 0.14\\
        centrals   & 1094 & 1.29 & 0.38 $\pm$ 0.04 & 0.98 & 3.53 $\pm$ 0.37\\
		\hline
	\end{tabular}
\end{table*}

\subsection{The cosmic HI mass density as a function of stellar mass}
\begin{figure}
    \centering
    \includegraphics[width=8cm]{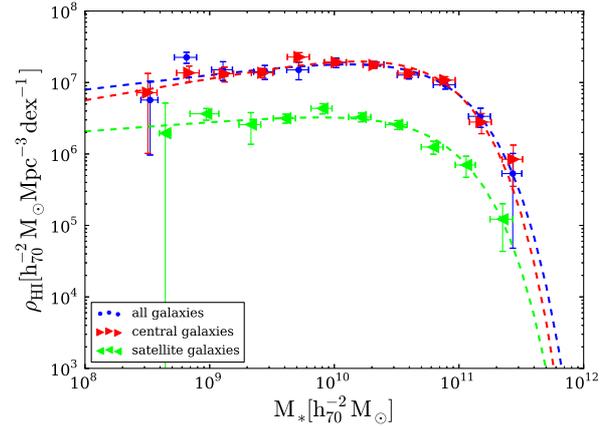}
    \caption{HI density as a function of stellar mass for the whole sample, centrals and satellites (same symbols as Fig.~\ref{stacking_stellarmassbins}). The HI density is obtained using the measured $\langle M_{\rm HI}\rangle$ from stacking and the stellar mass density from SDSS: $\rho_{\rm HI}(M_{\ast}) = \langle M_{\rm HI}(M_{\ast})\rangle\times\phi_{M_{\ast}}(M_{\ast})$. The dashed lines indicate the best Schechter fits to the data (see Table~\ref{density_function_parameters}).}
    \label{rho_HI_function_stellarmass}
\end{figure}

\begin{figure}
    \centering
    \includegraphics[width=8cm]{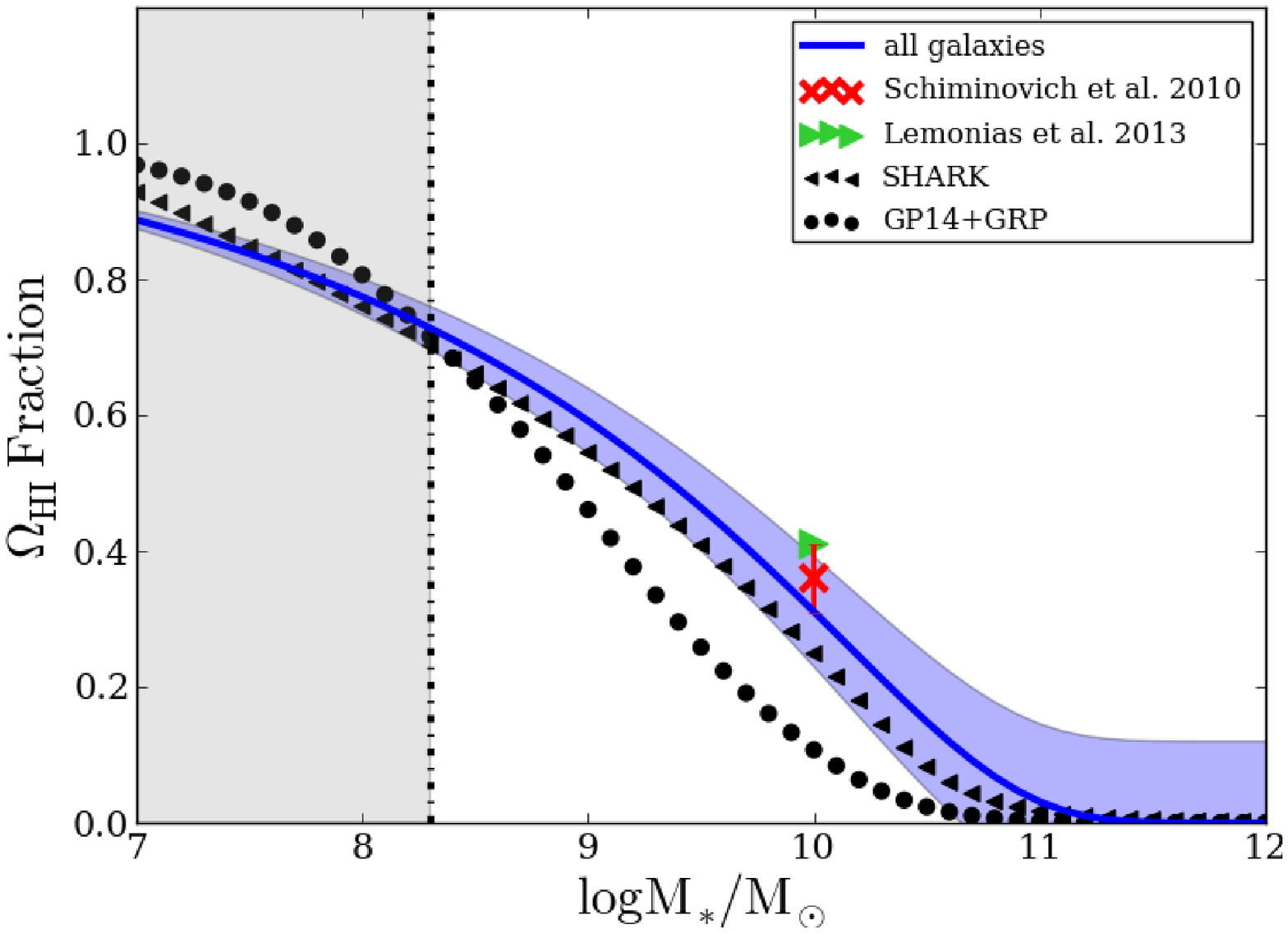}
    \includegraphics[width=8cm]{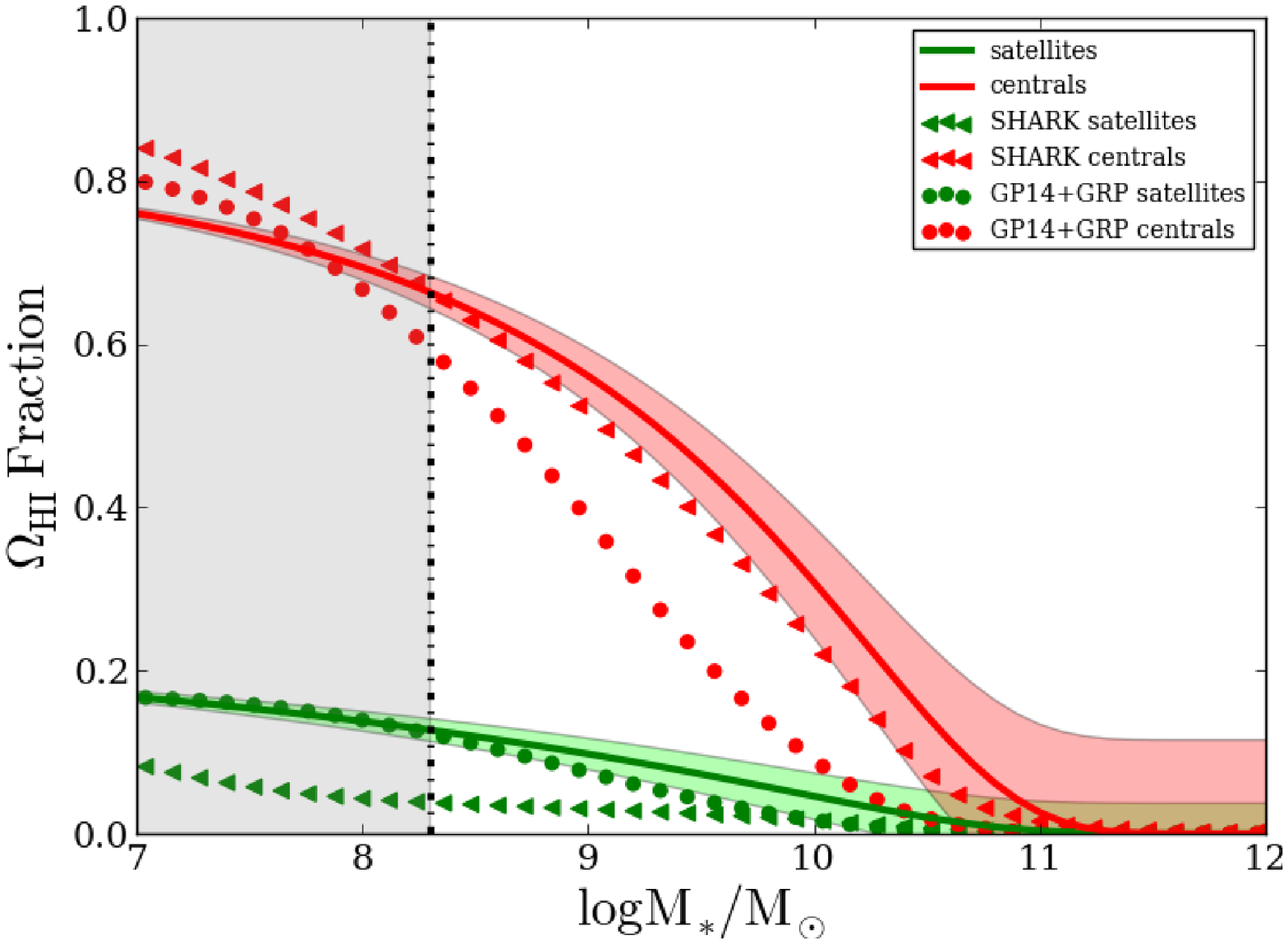}
    \caption{The cumulative fraction of $\Omega_{\rm HI}$ in galaxies above a given stellar mass for all galaxies (top panel, blue) and for satellites and centrals (bottom panel, green and red, respectively), with shaded regions showing the errors. In the top panel, we also present the $\Omega_{\rm HI}$ fraction measured by \citet[][red cross]{2010MNRAS.408..919S} and \citet[][green right triangle]{2013ApJ...776...74L}. In both panels, circles and left triangles show predictions from the GP14+GRP and {\sc Shark} semi-analytic models, respectively. The vertical dot-dashed line and light grey area indicate the region where our WSRT observations reached $M_{\rm HI}\sim$ 10$^{8.3}M_{\odot}$, below which the HI density is computed by extrapolation using the fitted $\rho_{\rm HI}(M_{\ast})$. This threshold happens to be similar to the resolution limit of the two simulations, $M_{*}\sim 10^8\,\rm M_{\odot}$.}
    \label{rho_HI_fraction_group}
\end{figure}

We can use the technique presented above to determine what is the contribution of galaxies of different stellar masses to the cosmic HI density of the local Universe, as this has been a matter of debate in the last few years. To do so, we integrate the best-fitting Schechter function to the $\rho_{\rm HI}-M_{\ast}$ relation shown in Fig.~\ref{rho_HI_function_stellarmass} (blue symbols and line) in different intervals of stellar mass. The results are presented as a cumulative distribution in Fig.~\ref{rho_HI_fraction_group} (top panel, blue line) and as differential bins in Table~\ref{omegahi_group_stellarmassbins_table}. We also present the corresponding results for the subsets of central and satellite galaxies in Fig.~\ref{rho_HI_function_stellarmass} (red and green symbols, respectively), Fig.~\ref{rho_HI_fraction_group} (bottom panel) and Table~\ref{omegahi_group_stellarmassbins_table}, For all the populations investigated in this paper, we find that galaxies below and above a stellar mass of $\sim10^{9.3}M_{\odot}$ contribute roughly equally to the total cosmic HI density. This mass threshold is slightly smaller than the stellar mass of M33 ($\sim$3-6 $\times$ 10$^{9}M_{\odot}$, \citealp{2003MNRAS.342..199C}) showing that, while low-mass galaxies are certainly important for the total HI mass budget of the local Universe, high-mass systems cannot be neglected -- indeed, $\sim$30\% of the atomic hydrogen in local galaxies is found in systems with stellar masses greater than $\sim10^{10}M_{\odot}$.

\begin{table}
 \caption{Parameters of the best Schechter fits to $\rho_{\rm HI}(M_{\ast})$ for all galaxies, satellites only and centrals only (see Fig.~\ref{rho_HI_function_stellarmass}).}
 \centering
 \label{density_function_parameters}
  \begin{tabular}{lccc}
   \hline
    Populations & $\phi^{\ast}_{M_{\ast}}$ & $\alpha$ & log$M^{\ast}$\\
     & ($10^{7}$Mpc$^{-3}$d$M_{\ast}$) & & (M$_{\odot}$)\\
   \hline
    All        & 3.07 $\pm$ 0.38 & $-$0.79 $\pm$ 0.04 & 10.79 $\pm$ 0.03\\
    satellites & 0.49 $\pm$ 0.08 & $-$0.86 $\pm$ 0.06 & 10.75 $\pm$ 0.04\\
    centrals   & 3.85 $\pm$ 0.48 & $-$0.69 $\pm$ 0.05 & 10.71 $\pm$ 0.04\\
   \hline
  \end{tabular}
\end{table}

\begin{table*}
 \caption{Measurements of $\Omega_{\rm HI}$ for the different galaxy populations in three stellar mass bins: $M_{\ast} \leqslant 10^{8}M_{\odot}$, $10^{8}M_{\odot} < M_{\ast} \leqslant 10^{10}M_{\odot}$, and $M_{\ast} > 10^{10}M_{\odot}$.}
 \label{omegahi_group_stellarmassbins_table}
  \begin{tabular}{lccc}
   \hline
    Populations & $\Omega_{\rm HI,M_{\ast}\leqslant 10^{8}M_{\odot}}$ &  $\Omega_{\rm HI,10^{8}M_{\odot} < M_{\ast}\leqslant 10^{10}M_{\odot}}$ & $\Omega_{\rm HI,M_{\ast} > 10^{10}M_{\odot}}$\\
     & $(10^{-4}h_{70}^{-1})$ & $(10^{-4}h_{70}^{-1})$ & $(10^{-4}h_{70}^{-1})$\\
   \hline
    All         & 0.90 $\pm$ 0.12 & 1.85 $\pm$ 0.24 & 1.24 $\pm$ 0.17\\
    satellites  & 0.29 $\pm$ 0.05 & 0.40 $\pm$ 0.07 & 0.20 $\pm$ 0.04\\
    centrals    & 0.50 $\pm$ 0.07 & 1.67 $\pm$ 0.22 & 1.34 $\pm$ 0.20\\
   \hline
  \end{tabular}
\end{table*}

Interestingly, \citet{2010MNRAS.408..919S} used $\sim$ 190 galaxies from the GASS survey (M$_{\ast} > 10^{10}$ M$_{\odot}$, 0.025 < z < 0.050) and found that 36 $\pm$ 5 per cent of the total HI mass density is in galaxies with M$_{\ast} > 10^{10}$ M$_{\odot}$.  More recently, using a sample of 480 galaxies from the second data release of GASS, 
\citet{2013ApJ...776...74L} computed the bivariate HI mass-stellar mass function for the range of stellar masses targeted by GASS, $\Omega_{\rm HI,M_{\ast}>10^{10}M_{\odot}}$, and found that massive galaxies contribute 41$\%$ of the HI density in the local Universe. We plot these two values in the top panel of Fig.~\ref{rho_HI_fraction_group} as a red cross and a green triangle, respectively. Our findings are consistent, within errors, with both estimates, with a slightly better agreement with \citet{2010MNRAS.408..919S}. This is remarkable given that both the samples and the techniques used in this work are very different from the ones used by \citet{2010MNRAS.408..919S} and \citet{2013ApJ...776...74L}, providing independent support to the important contribution of massive galaxies to the cosmic HI mass budget of the local Universe. 


\section{Comparison with semi-analytical models of galaxy formation}
\label{sec:simulation}

In this section we show that our measurements offer stringent constraints to galaxy formation simulations, by  providing a clear separation between the contributions of centrals and satellite galaxies of different stellar masses to $\Omega_{\rm HI}$. 


\citet{2014MNRAS.440..920L} presented predictions for the contribution of galaxies with different stellar masses to the cosmic densities of atomic and molecular hydrogen in the context of galaxy formation in a $\Lambda$CDM framework. They use three flavours of the semi-analytic model (SAM) of galaxy formation GALFORM \citep{2000MNRAS.319..168C}: the Lagos12 \citep{2013MNRAS.436.1787L}, Gonzalez-Perez14 \citep{2014MNRAS.439..264G} and Lacey16 \citep{2016MNRAS.462.3854L} models. In these three models, they found the density of HI to be always dominated by galaxies with low stellar masses ($M_{\ast} < 10^9$ M$_{\odot}$), clearly in tension with our findings. 

In order to perform a more accurate comparison between our findings and those of \citet{2014MNRAS.440..920L}, we focus on the  $z=0$ simulated $(500\,\rm Mpc/h)^3$ box of GALFORM \citep{2014MNRAS.439..264G}, which includes a treatment of gradual ram-pressure stripping for satellite galaxies described in \citet{2014MNRAS.443.1002L}. We refer to this model as ``GP14+GRP'', following the naming convention adopted by the authors. The gradual ram-pressure stripping implementation allows satellite galaxies to continue to experience gas accretion after they cross the virial radius of the group, thus increasing the timescale needed for the quenching of the star formation. From the model, we select all galaxies with stellar mass $M_{\ast} >10^5\,\rm M_{\odot}$ and calculate $\Omega_{\rm HI}$ using:
\begin{eqnarray}
     \Omega_{\rm HI} = \Sigma_{i}{M^{i}_{\rm HI}}/V,
     \label{omega_HI_simulation}
\end{eqnarray}
where $M^{i}_{\rm HI}$ is the HI mass of the $i$-th galaxy and V is the total simulated volume. However, it is worth mentioning that the resolution of this model translates into a stellar mass limit of $\sim 10^{8}\,\rm M_{\odot}$, below which galaxies are not expected to be converged. This roughly corresponds to the stellar mass limit of our sample, as indicated by the grey area in Fig.~\ref{rho_HI_fraction_group}. The cumulative distributions for $\Omega_{\rm HI}$ as a function of stellar mass obtained for all galaxies in GP14+GRP and for centrals and satellites separately are shown by the filled circles in Fig.~\ref{rho_HI_fraction_group}. As expected, the GP14+GRP model significantly underestimates the contribution of galaxies with stellar masses greater than 10$^{9}$ M$_{\odot}$ to the atomic gas mass density in the local Universe. Interestingly, despite the well known limitation of the GP14+GRP model in reproducing the properties of satellite galaxies \citep{2017MNRAS.466.1275B}, the mismatch that we see in this case is mainly driven by central galaxies. For galaxies more massive than $\sim$10$^{9.5}$ M$_{\odot}$, the model predicts a factor of $\sim$2 less gas than what observed in our sample.    


Recently, \citet{2018MNRAS.481.3573L} presented a new semi-analytic model of galaxy formation, {\sc Shark}, with improvements over previous SAMs in the ability to reproduce galaxy scaling relations. \citet{Chauhan2019} also recently showed that {\sc Shark} is able to reproduce the HI mass-velocity width relation observed by the ALFALFA survey. It is thus interesting to see if the tension between observations and simulations extends to the {\sc Shark} implementation as well. 

We use the {\sc Shark} $z=0$ simulated box of $(210\,\rm Mpc/h)^3$ volume. This model assumes instantaneous stripping of the hot gas of satellites, which means that their hot halo is stripped as soon as they cross the virial radius of their group. This generally causes a relatively fast exhaustion of their interstellar medium and star formation quenching. We analyse {\sc Shark} exactly in the same way as the GP14+GRP model and the results are shown in Fig.~\ref{rho_HI_fraction_group} as triangles. 

 


We find that {\sc Shark} is in much better agreement with our observations for central galaxies. For example, about 25$\%$ of the cosmic HI density is located in galaxies with $M_{\ast} > 10^{10} M_{\odot}$ at redshift $z=0$, while the GP14+GRP model predicts a contribution of $\sim$ 9\%.  
The situation is, however, reversed for satellite galaxies, where the GP14+GRP model gives a better match to the observations than {\sc Shark}. The latter is likely driven by the different treatment of hot halo stripping of satellites in the models, in which the former applies a gradual stripping, while the latter assumes instantaneous stripping. 


Conversely, it is less obvious why {\sc Shark} more closely matches the observations for the full sample and centrals alone than the GP14+GRP model. A closer comparison between the predictions of the two models suggests that the difference most likely lies in the way the partition between atomic and molecular hydrogen is set. While the two models predict very similar total cold gas masses at fixed stellar mass, above stellar masses of 10$^{10}$ M$_{\odot}$ the fraction of cold gas mass in {\it atomic} form is $\sim$10 times higher in {\sc Shark} than in the GP14+GRP model.

This is intriguing, as both models assume a pressure-HI/H$_2$ relation based on the empirical model of \citet{Blitz2006}. In this model, $\Sigma_{\rm H_2}/\Sigma_{\rm HI}=(P/P_{0})^{\alpha}$, with $\Sigma_{\rm H_2}$ and $\Sigma_{\rm HI}$ being the surface densities of molecular and atomic hydrogen, respectively, $P$ being the hydrostatic pressure, and $P_{0}$ and $\alpha$ being observationally constrained. However, {\sc Shark} adopts the \citet{Blitz2006} reported value of $P_{0}/k_{\rm B}=34,273 \,\rm K\,cm^{-3}$ (with $k_{\rm B}$ being Boltzmann's constant), which is about $2$ times larger than the value adopted in the GP14+GRP model, $P_{0}/k_{\rm B}=17,000\,\rm K\, cm^{-3}$, which is based on \citet{Leroy2008}. This effectively makes the atomic-to-molecular conversion less efficient in {\sc Shark} compared to GALFORM, which allows galaxies to be more HI-rich for the same star formation rate. However, it is worth emphasizing that the atomic-to-molecular conversion efficiency in both models also depends on the accretion plus feedback cycle (which are different), and hence, the different $P_{0}$ values are likely only partially responsible for the differences seen in the models.

Although we cannot conclusively point to the main physical process responsible for the difference seen in central galaxies between the two simulations discussed here, it is clear that they would overall greatly benefit from using gas observations, such as those shown here,  to constrain their free parameters. This is because both models broadly reproduce other measurements, such as the stellar mass function and the star-formation rate-stellar mass relation, 
showing that the gas content of galaxies, and particularly the contribution of centrals/satellites as a function of stellar mass, provides a strong, independent constraint. 



\section{Discussion \& Conclusion}
\label{sec:discussion}
In this paper we use an interferometric stacking technique to study the contribution of centrals and satellites of different stellar masses to the cosmic HI mass density in the local Universe.


We show that, as expected, $\Omega_{\rm HI}$ is dominated by central galaxies at the mean redshift of $\langle z\rangle = 0.065$. We then present, for the first time, the distribution of $\Omega_{\rm HI}$ in stellar masses for galaxies in different hierarchies and find that galaxies with stellar masses above 10$^{10}$ M$_{\odot}$  contribute to $\sim$30\% of the total atomic hydrogen in local galaxies, and that 50\% of $\Omega_{\rm HI}$ is reached around stellar masses of $\sim$10$^{9.3}$ M$_{\odot}$. 

While our findings are consistent with previous determinations of the contribution of massive galaxies to $\Omega_{\rm HI}$ \citep{2010MNRAS.408..919S,2013ApJ...776...74L}, they are in tension with \citet{2014MNRAS.440..920L}, who showed that in semi-analytic models most of the HI is stored in galaxies with masses below $\sim$10$^{9}$ M$_{\odot}$ and, most importantly, that massive galaxies ($M_{\ast} >10^{10}$ M$_{\odot}$) contribute to only 9\% of $\Omega_{\rm HI}$. In order to fully understand the origin of this tension, we extend the work by \citet{2014MNRAS.440..920L} and compare our results with predictions from the semi-analytic models 
GP14+GRP and {\sc Shark}. 

In the case of the whole sample or central galaxies only, {\sc Shark} more closely matches our findings.  We show that this is - at least partially - due to the different prescriptions used for the partition between atomic and molecular hydrogen in the two models. In GP14+GRP massive galaxies appear too atomic hydrogen poor than observed. This is intriguing, as it would also mean that the contribution of massive galaxies to $\Omega_{\rm H2}$ found by \citet{2014MNRAS.440..920L} might be in reality significantly smaller, and that the overlap in stellar mass between the galaxy populations dominating $\Omega_{\rm HI}$ and $\Omega_{\rm H2}$ is actually wider than previously claimed. Of course, this cannot be confirmed until a similar analysis for the molecular hydrogen content of galaxies is performed. 

Conversely, when it comes to satellite galaxies, the GP14+GRP implementation of environmental effects produces results closer to ours than {\sc Shark}. This implies that the observed $\Omega_{\rm HI}$ distribution in stellar mass for satellites cannot be reproduced by strangulation of the gas alone and gradual ram-pressure stripping works better. This is fully consistent with the recent work by \citet{2017MNRAS.466.1275B}, who compared the results from stacking of 10,600 satellite galaxies extracted from the ALFALFA survey footprint with both SAMs (GP14 and GP14+GRP) and hydrodynamical simulations \citep{2013MNRAS.434.2645D}. However, it is promising to note that hydrodynamical models are quickly improving and that some of the tensions highlighted by \citet{2017MNRAS.466.1275B} are being addressed \citep[e.g.,][]{Stevens2019}. 



In conclusion, our work highlights how a simple parametrisation of $\Omega_{\rm HI}$ as a function of stellar mass and hierarchy can still bring to the surface important limitations in our current understanding of the gas cycle in galaxies, and provide fundamental constraints to cosmological simulations. Thanks to the advent of the Square Kilometer pathfinder telescopes such as Australian Square Kilometre Array Pathfinder (ASKAP) \citep{2008ExA....22..151J,2009pra..confE..15M}, MeerKAT \citep{2012IAUS..284..496H}, Five-hundred-meter Aperture Spherical radio Telescope (FAST) \citep{2011IJMPD..20..989N,2008MNRAS.383..150D} and WSRT/Aperture Tile in Focus (APERTIF) \citep{2009wska.confE..70O}, it will very soon be possible to extend this approach to significantly larger samples making it possible to further dissect the contribution of different galaxy populations to the atomic gas mass density of the local Universe, and potentially extend this to higher redshifts.

\label{sec:summary}

\section{Acknowledgements}

The WSRT is operated by ASTRON (Netherlands Foundation for Research in Astronomy) with support from the Netherlands Foundation for Scientific Research (NWO). This research made use of the `K-corrections calculator' service available at http://kcor.sai.msu.ru/. We acknowledge the use of Miriad software in our data analysis
(http://www.atnf.csiro.au/computing/software/miriad/). This research made use of the Sloan Digital Sky Survey archive. The full acknowledgment can be found at http://www.sdss.org. Parts of this research were supported by the Australian Research Council Centre of Excellence for All Sky Astrophysics in 3 Dimensions (ASTRO 3D), through project number CE170100013. LC is the recipient of an Australian Research Council Future Fellowship (FT180100066) funded by the Australian Government.

\bibliographystyle{mnras}
\bibliography{group}

\bsp	
\label{lastpage}
\end{document}